\documentclass[twocolumn,showpacs]{revtex4}
\usepackage{graphicx}
\usepackage{natbib}
\usepackage{latexsym}
\newcommand{\bfr}{{\bf r}}
\newcommand{\bfR}{{\bf R}}

\newcommand{\hpsi}{\hat{\psi}}

\newcommand{\hPsi}{\hat{\Psi}}

\newcommand{\ddr}{\frac{\partial}{\partial r}}

\begin{document}
\title{A pseudo-potential analog for zero-range photoassociation and Feshbach resonance}
\author{M. G. Moore}
\affiliation{Department of Physics and Astronomy, Ohio University, Athens, OH 45701}
\date{\today }
\begin{abstract}
A zero-range approach to atom-molecule coupling is developed in analogy to the Fermi-Huang pseudo-potential treatment of atom-atom interactions. It is shown by explicit comparison to an exactly-solvable finite-range model that replacing the molecular bound-state wavefunction with a regularized delta-function can reproduce the exact scattering amplitude in the long-wavelength limit. Using this approach we find an analytical solution to the two-channel Feshbach resonance problem for two atoms in a spherical harmonic trap. 
\end{abstract}
\maketitle

Coupling between atoms and molecules in quantum-degenerate gases is an ever-present aspect of ultracold atomic physics. Feshbach resonances (FR)  \cite{InoAndSte98} are now routinely used for control over atomic interactions \cite{CorClaRob00} and the formation of molecular Bose-Einstein condensates \cite{JocBarAlt03,GreRegJin03,XuMukAbo03,DurVolMar04}. Laser-induced photoassociation (PA) is also widely employed \cite{TheThaWin04,WynFreHan00}, having the advantage of control over the coupling strength \cite{ProPicJun03}. While a  zero-range approach to atom-atom collisions has long been a cornerstone of BEC theory, an analog to the Fermi-Huang pseudo-potential approach \cite{FerHua} has yet to be formulated to treat multichannel free-bound coupling in ultra-cold atomic gases.

In the long-wavelength limit, the energy-dependence of the scattering phase-shift for atomic collisions takes a universal form, with all information about the details of the interaction potential contained in a single parameter, the scattering length. As a result, the full interaction potential can be replaced by a regularized delta-function pseudo-potential, which yields the correct scattering amplitude up to a third-order correction in the ratio of the effective range to the incident wavelength. In this Letter we formulate an analogous approach to atom-molecule coupling, by replacing the bound-state wavefunction with the zero-range object which correctly reproduces the long-wavelength scattering amplitude. The resulting model contains no divergences and does not require a momentum cut-off. It is likely that this model will play an important role in understanding the role played by atom-atom correlations in FR and PA physics, particularly in the strong-coupling regime, where such effects play a dominant role.

The first zero-range model for BEC atom-molecule coupling, proposed by Heinzen and coworkers, replaced the bound-state wavefunction with a delta-function \cite{HenWynDru00}. This approach was shown by Holland and coworkers to contain a UV divergence when pair correlations were taken into account \cite{HolParWal00}, thus limiting its applicability. Holland and coworkers demonstrated that this divergence could be removed via a momentum cut-off and re-normalized detuning. As we will see, this approach fails in the presence of a  background scattering length. 

We begin our analysis by considering a pair of atoms described by a relative wavefunction $\phi_j(\bfr,t)$, where $j=1,2$ corresponds to an internal spin state. The eigenstates of this system obey the Schr\"odinger equation,
\begin{equation}
\label{schro}
	E\phi_j(\bfr)=-\frac{\hbar^2}{2\mu}\nabla^2\phi_j(\bfr)+\sum_{k}V_{jk}(\bfr,t)\phi_k(\bfr),
\end{equation}
where $E$ is the energy eigenvalue, $\mu$ is the reduced mass and $V_{jk}(\bfr)$ is the inter-atom potential. For our model system we assume that the first channel sees a flat potential, $V_{11}(\bfr)=0$. The second channel sees a spherical-well potential of depth $V_0$ and radius $w$, $V_{22}(\bfr)=U_0-V _0U(w-r)$, where $U_0$ is the continuum threshold energy and $U(x)$ is the unit-step function. In the absence of coupling terms, i.e. for $V_{12}(\bfr)=0$, the spectrum of the second channel consists of a continuum of states above the threshold energy, $U_0$, and a discrete set of bound states with energies between $U_0$ and $U_0-V_0$. The bound-states are all of the form
\begin{equation}
\label{phib}
	\psi_b(\bfr)=\left\{ \begin{array}{cc} 
	{\cal N}_b \frac{e^{-r/a_b}}{r}& :r >w \\	
	{\cal N}_b\frac{e^{-w/a_b}}{\sin(k_bw)}\frac{\sin(k_br)}{r}&: r <w\\
	\end{array}\right.
\end{equation}
where $a_b$ and $k_b$ satisfy the equations $\frac{\hbar^2}{2\mu}\left[k_b^2+1/a_b^2\right]=V_0$ and $\cot(k_bw)=-1/(k_ba_b)$, and ${\cal N}_b$ is determined by normalization. The bound state energies are $E_b=U_0-\hbar^2/(2\mu a_b^2)$.

We proceed by first expanding the second channel wavefunction, $\phi_2(\bfr)$, onto its bare eigenstates under the simplifying assumptions that only a single bound state is near-resonantly coupled to the first channel so that all other states may be neglected. We assume the interaction potential has the form $V_{12}(\bfr,t)\approx\frac{\hbar^2G}{\mu}e^{-i\omega t}$. Taking $E=\hbar^2k^2/(2\mu)$ then leads to an eigenvalue problem for a continuum coupled to a single bound state,
\begin{eqnarray}
\label{amscatt1}
	\frac{1}{2}\left[k^2+\nabla^2\right]\phi_1(\bfr)&=&G \psi_b(\bfr) c\\
\label{amscatt2}
	\frac{1}{2}\left[k^2-2\Delta\right]c&=&G^\ast\int d^3r\, \psi^\ast_b(\bfr) \phi_1(\bfr),
\end{eqnarray}
where $c$ is the probability amplitude for the atom pair to be in the bound state, and $\Delta=\frac{\mu}{\hbar^2}(U_0-\omega)-\frac{1}{2a_b^2}$ is the detuning away from the atom-molecule resonance at $k=0$. The coupling constant $G$ will depend on the details of the atom-molecule coupling scheme. 

Our goal is now to solve this eigenvalue problem, under the boundary conditions 
\begin{eqnarray}
\label{bc1}
	\lim_{r\to\infty}\phi_1(\bfr)&=&\frac{e^{-ikr}}{r}+f\frac{e^{ikr}}{r}\\
\label{bc2}
	\lim_{r\to 0}r\phi_1(\bfr)&=&0,
\end{eqnarray}
in order to determine the scattering amplitude $f=f(k)$. The solution can be obtained via the ansatz
\begin{equation}
\label{ansatz}
	\phi_1(\bfr)=\left\{ \begin{array}{cc}
	\frac{e^{-ikr}}{r}+f\frac{e^{ikr}}{r}+\frac{2G a_b^2 }{1+(a_bk)^2}c\, \psi_b(\bfr) &:r>w\\
	\beta \frac{\sin(kr)}{r}+\frac{2G}{k^2-K_b^2}c\, \psi_b(\bfr) &:r<w\\
	\end{array}\right. .
\end{equation}
This ansatz explicitly satisfies (\ref{amscatt1}), as well as the boundary conditions (\ref{bc1}-\ref{bc2}). Equation (\ref{amscatt2}), together with the continuity equations $\phi_1(w^+)=\phi_1(w^-)$ and $\nabla\phi_1(w^+)=\nabla\phi_1(w^-)$ can then be used to determine the three unknowns $f$, $c$, and $\beta$. These equations are linear in the three unknowns, and can be thus solved in a straightforward manner. 

The long-wavelength limit requires that $1/k$ be large compared to the size of the bound-state. As the size of the bound-state is $w+a_b$, this is equivalent to the limits $kw\ll 1$ and $ka_b\ll 1$. For our model potential the condition $K_b>1/w$ is always satisfied, so that $k/K_b$ is a small parameter as well. Expanding the scattering amplitude $f(k)$ in terms of these small parameters then yields
\begin{equation}
\label{f}
	f(k)=-\frac{k^2-2\delta-ik\frac{|\chi|^2}{\pi}}{k^2-2\delta+ik\frac{|\chi|^2}{\pi}}+O[\varepsilon^3],
\end{equation}
where $\varepsilon\in\{kw,ka_b,k/K_b\}$, and we
 have introduced the light-shifted detuning
\begin{eqnarray}
\label{delta}
	\delta&=&\Delta-8\pi|G|^2{\cal N}_b^2e^{-2w/a}
	\left[\frac{e^{2w/a}}{{\cal N}_b^2K_b^2}-\left[1+\frac{1}{K_b^2a_b^2}\right]\frac{a_b^3}{2}\right. \nonumber\\
	& &+\left. \left[1+\frac{1}{K_b^2a_b^2}\right]^2a_b^2(a_b+w)\right],
\end{eqnarray}
and the effective coupling constant
\begin{equation}
\label{chi}
	\chi=4\pi G{\cal N}_be^{-w/a}\,a_b(a_b+w)\left[1+\frac{1}{K_b^2a_b^2}\right].
\end{equation}
The important point here is that all of the details of the potential can be absorbed into effective detuning and coupling constants. 

We now consider a zero-range model in which the bound-state wavefunction $\psi_b(\bfr)$ in (\ref{amscatt1}) is replaced by a regularized delta-function, $G\psi_b(\bfr)\to\chi\delta^3(\bfr)\ddr r$. In addition, the detuning $\Delta$ is replaced by the light-shifted detuning $\delta$ and the coupling constant $G$ is replaced by the effective coupling constant $\chi$. The Schr\"odinger equation for this model is given by
\begin{eqnarray}
\label{zr1}
	\frac{1}{2}\left[k^2+\nabla^2\right]\phi_1(\bfr)&=&\chi \delta^3(\bfr) c\\
\label{zr2}
	\frac{1}{2}\left[k^2-2\delta\right]c&=& \chi^\ast\int d^3r\, \delta^2(\bfr)\ddr r\phi_1(\bfr).
\end{eqnarray}
This problem can be solved by making use of  the ansazt $\phi_1(\bfr)=\frac{e^{-ikr}}{r}+f\frac{e^{ikr}}{r}$ and the identity $\nabla^2\frac{1}{r}=-4\pi\delta^3(\bfr)$. The scattering amplitude is readily found to be
\begin{equation}
\label{fzr}
	f(k)=-\frac{k^2-2\delta-ik\frac{|\chi|^2}{\pi}}{k^2-2\delta+ik\frac{|\chi|^2}{\pi}},
\end{equation}
which agrees with the result (\ref{f}) up to a correction of third-order in the small parameters $kw$, $ka_b$, and $k/K_b$. Thus the zero-range model (\ref{zr1}-\ref{zr2}) will reproduce correctly the long-wavelength atom-molecule quantum-dynamics of our model potential. 

Second quantization of this model yields the Hamiltonian
\begin{eqnarray}
\label{H}
	& &\hat{\cal H}=-\frac{\hbar^2}{4m}\int d^3r\left[\hpsi^\dag(\bfr)\nabla^2\hpsi(\bfr)
	+\frac{1}{2}\hPsi^\dag(\bfr)\nabla^2\hPsi(\bfr)\right]\nonumber\\
	& &+\frac{\hbar^2\chi}{\sqrt{2}m}\int d^3R\, d^3r\, \hPsi^\dag(\bfR)\delta^3(\bfr)\ddr r\, \hpsi(\bfR+\frac{\bfr}{2})\hpsi(\bfR-
	\frac{\bfr}{2}) \nonumber\\
	& &+H.c.
\end{eqnarray}
where $\hpsi(\bfr)$ is the annihilation  operator for an atom of mass $m=2\mu$, and $\hPsi(\bfr)$ is the annihilation operator for a molecule of mass $2m$. The system of equations (\ref{zr1}-\ref{zr2}) can be derived from this Hamiltonian via the 2-atom quantum state 
\begin{eqnarray}
\label{Psi}
	|\Psi\rangle&=&\frac{1}{\sqrt{2}}\int d^3Rd^3r\, \Phi(\bfR)\phi(\bfr)\hpsi^\dag(\bfR+\frac{\bfr}{2})
	\hpsi^\dag(\bfR-\frac{\bfr}{2})|0\rangle\nonumber\\
	&+&c\int d^3R\Phi(\bfR)\hPsi^\dag(\bfR)|0\rangle,
\end{eqnarray} 
where $\Phi(\bfR)$ is an arbitrary center-of-mass wavefunction and $|0\rangle$ is the vacuum state.
The Hamiltonian (\ref{H}) should form the basis of any field-theoretical description of zero-range atom-molecule coupling.

As an example, we now solve the problem of two bosonic atoms in a spherical harmonic oscillator (with frequency $\omega_{trap}$) with both s-wave collisions and coupling to a bound state in a second channel. With $E=\hbar\omega_{trap}(\nu_n+3/2)$, $\delta\to\hbar\omega_{trap}\delta$, and using harmonic oscillator units, the time-independent Schr\"odinger equation can be written as
\begin{eqnarray}
\label{3dshofr1}
	&\left[\nu_n+\frac{1}{2}\nabla^2-\frac{1}{2}r^2+\frac{3}{2}\right]\phi_n(\bfr)=
	2\pi\left(\frac{a}{\lambda}\right)\delta^3(\bfr)\ddr r\, \phi_n(\bfr)&\nonumber\\
	&+\pi^{3/4}\Omega\delta^3(\bfr)\, c_n&\\
\label{3dshofr2}
	&\left[\nu_n-\delta\right]c_n=\pi^{3/4}\Omega\int d^3r\, \delta^3(\bfr)\ddr r\, \phi_n(\bfr),&
\end{eqnarray}
where $n$ is an integer label for each quantum level (the lowest energy level corresponding to $n=0$), $a$ is the background scattering length, $\lambda$ is the harmonic oscillator length of the trap, and $\Omega=\lambda^2\pi^{-3/4}\chi$.
The normalized eigenfunctions are found to be \cite{BusEngRza98} 
\begin{equation}
\label{phin}
	\phi_n(\bfr)=-\frac{\Omega}{2\pi^{3/4}} \frac{\Gamma[-\frac{\nu_n}{2}]}{\beta(\nu_n,\frac{a}{\lambda})} c_n e^{-r^2/2}
	U(-\frac{\nu_n}{2},\frac{3}{2},r^2),
\end{equation}
\begin{equation}
\label{cn}
	c_n=\left[1+\frac{\Omega^2\sqrt{\pi}}{2}\frac{\Gamma[-\frac{\nu_n}{2}]}{\Gamma[-\frac{\nu_n+1}{2}]}\frac{[\psi(-\frac{\nu_n}{2})-\psi(-\frac{\nu_n+1}{2})]}{\beta^2(\nu_n,\frac{a}{\lambda}) }\right]^{-1/2},
\end{equation}
where $\beta(\nu,x)=1-2x \Gamma[-\frac{\nu}{2}]/\Gamma[-\frac{\nu+1}{2}]$, 
 $U(a,b,z)$ is the confluent hypergeometric function and $\psi(z)$ is the polygamma function \cite{AbrSte65}. The eigenvalues $\{ \nu_n\}$ are determined by the characteristic equation
\begin{equation}
\label{CE}
	\delta=\nu_n-\frac{\Omega^2\sqrt{\pi}\Gamma[-\frac{\nu_n}{2}]}{\Gamma[-\frac{\nu_n+1}{2}]\beta(\nu_n,\frac{a}{\lambda})},
\end{equation}
where there is an apparently non-trivial relation $|c_n|^2=d\nu_n/d\delta$. It is straightforward to show that the spectrum of eigenvalues will agree exactly with those of a single-channel system with the energy-dependent effective scattering length
\begin{equation}
\label{aeff}
	a_{eff}(\nu)=a+\frac{\lambda}{2}\frac{\sqrt{\pi}\Omega^2}{(\nu-\delta)},
\end{equation}
which is the familiar Feshbach Resonance result. The only difference between the true atom-molecule eigenstates and the equivalent single-channel states with scattering length $a_{eff}$, is the presence of the bare-molecule population, $|c_n|^2$. From a series expansion of (\ref{phin}) the $1/r$ part of $\phi_n(\bfr)$ is found to be $-\frac{\Omega}{2\pi^{1/4}\beta(\nu_n,\frac{a}{\lambda})}\frac{c_n}{r}$. Only for $a=0$ is this term independent of $\nu_n$, so that it can be removed via a renormalized detuning \cite{HolParWal00}.

On resonance we have $\nu_n=\delta$ and $|a_{eff}|\to\infty$. A careful analysis shows that this requires $\nu_n=2n-1$ and $c_n\neq 0$. Thus the eigenvalues are driven to  odd-integer or 'fermionized' values, for which the regular part of $\phi_n(\bfr)$ vanishes at $\bfr=0$. Inserting this result into Eq. (\ref{cn}) gives an analytic expression for the on-resonance molecular fraction,
\begin{equation}
\label{cres}
	|c_n|^2=\frac{1}{1+\alpha_n\Omega^2},
\end{equation}
where $\alpha_n=\frac{(2n)!!}{(2n-1)!!}\pi/2$. For the low lying levels we have $\alpha_0=\pi/2$, $\alpha_1=\pi$ and $\alpha_2=4\pi/3$. 
\begin{figure}
\includegraphics[scale=.85]{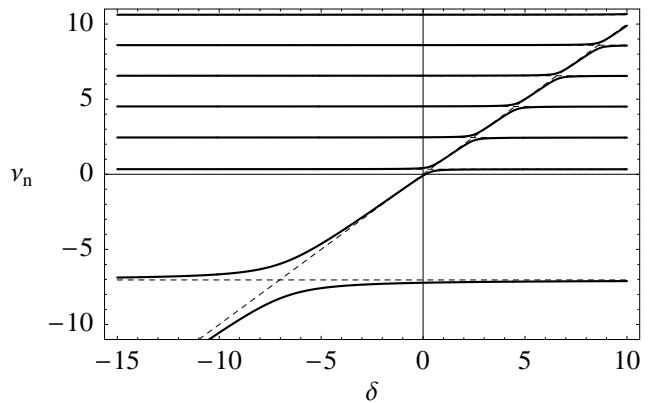}
\caption{\label{fig1} The eigenvalue spectrum as a function of the detuning for the case $a=.3\lambda$ and $\Omega=.2$, illustrating a sequence of avoided crossings in the weak-coupling regime. The dashed lines correspond to the uncoupled eigenvalues}
\end{figure}
\begin{figure}
\includegraphics[scale=.85]{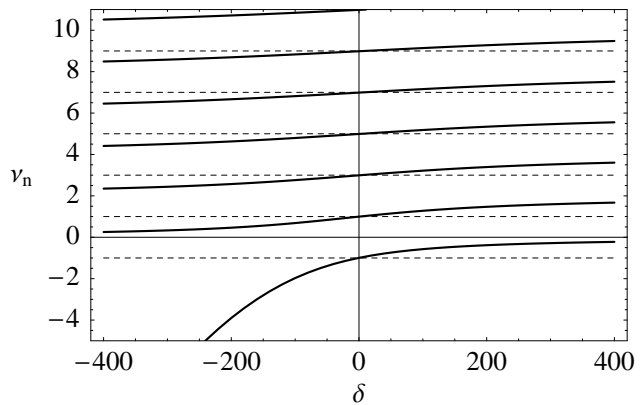}
\caption{\label{fig2} The eigenvalue spectrum as a function of the detuning for the case $a=0$ and $\Omega=10$, illustrating the 'fermionization' of the low lying levels in strong-coupling regime (in the vicinity of $\delta=0$). The dashed lines correspond to the odd-integer values $\nu_n=2n-1$.}
\end{figure}
\begin{figure}
\includegraphics[scale=.85]{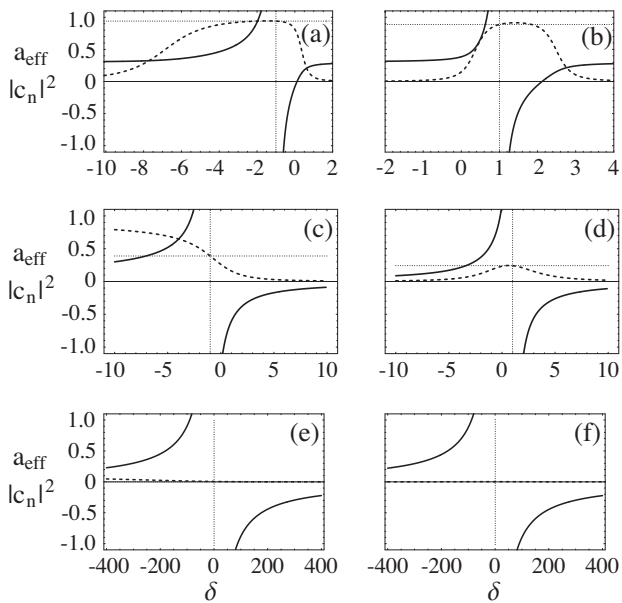}
\caption{\label{fig3} Effective scattering length, $a_{eff}$, (solid line) and molecular fraction, $|c_n|^2$, (dashed line) as the detuning, $\delta$, is swept across resonance. Figures 3a and 3b show the cases $n=1$ and $n=2$, respectively, for the case $\Omega=.2$ and $a=.3$. Figures 3c and 3d show $n=0$ and $n=1$ for $\Omega=1$ and $a=0$, while figures 3e and 3f show $n=0$ and $n=1$ for the case $\Omega=10$ and $a=0$. The vertical dotted lines mark the location of the resonance, while the horizontal dotted lines correspond to the analytical result for $|c_n|^2$ given by Eq.  (\ref{cres}).}
\end{figure}

The energy-dependence in the effective scattering length is critical to understanding the cross-over between the weak-coupling and strong coupling regimes. The requirement for a significant deviation from the bare-trap spectrum is $a_{eff}/\lambda\sim1$. Obtaining this condition via Feshbach resonance requires $\delta=\tilde{\nu}_n\pm\sqrt{\pi}\Omega^2/2$. If this width is smaller than the level spacing, only a single level can be near-resonant for a given detuning. In this weak-coupling regime, $\Omega^2\ll 1$, the spectrum consists of a series of avoided crossings between the bare molecular level and the uncoupled eigenstates of the 'open' channel. At each avoided crossing there will be strong mixing between a single trap level and the molecular state. Sweeping the detuning can select which trap level is resonantly coupled to the molecular state. This is illustrated in Figure \ref{fig1}, where we have plotted the eigenvalue spectrum as a function of the detuning for the case $a=.3\lambda$ and $\Omega=.2$. The dotted lines show the uncoupled ($\Omega=0$) eigenvalues. The shifts in the asymptotic values of the energy levels from the bare trap spectrum ($\nu_n=2n$) are due to the presence of s-wave collisions. The asymptotic state at $\approx-.7$ is the bound state of the 'open' channel, which is an eigenstate of the trap plus pseudo-potential system.

In the strong coupling regime, defined as $\Omega^2\gg1$, the width of the resonance is much larger than the trap level-spacing, hence many levels can be resonant simultaneously. Thus the low-lying levels all lie very close to their on-resonance values of $\nu_n=2n-1$. This is illustrated in Figure \ref{fig2}, which shows the eigenvalue spectrum as a function of detuning for the case $\Omega=10$ and $a=0$. In this regime Eq. (\ref{cres}) is a good estimate for the molecular fraction, showing that the molecular amplitude decreases dramatically with increasing coupling strength. To understand this effect, we simply make the reasonable assumption that in the strong-coupling limit all quasi-resonant levels are mixed with equal amplitudes. For $\Omega^2\gg 1$, the number of near resonant levels is $N_{levels}\approx \Omega^2$. If we equate the probability for any given bare-state to the  total probability divided by the approximate number of levels we arrive at $|c|^2\approx1/\Omega^2$, which agrees well with Eq. (\ref{cres}).

In Figure 3 we plot $a_{eff}$ and $|c_n|^2$ versus detuning for several cases of interest. In Figs 3a and 3b we show the weak-coupling case $\Omega=.2$ and $a=.3$ for levels $n=1$ and $n=2$ respectively. The $n=1$ case shows a sweep (right to left) from the lowest `unbound' state into the bound state in the `open' channel. The $n=2$ case shows a transfer from one `unbound' state to another. As the level is swept through resonance we see a broad feature in the molecular fraction $|c_n|^2$, whose maximum value is slightly larger than the on-resonance value (\ref{cres}) and occurs to the right of the resonance. Figures 3c and 3d show the intermediate case $\Omega=1$ and $a=0$ for levels $n=0$ and $n=1$. We see in the $n=1$ case that the molecular fraction is significantly reduced compared to the weak-coupling regime. Lastly, in Figures 3e and 3f we see the strong-coupling case $\Omega=10$ and $a=0$, for levels $n=0$ and $n=1$. We see that in the strong coupling regime, the scattering length can be tuned from $-\infty$ to $+\infty$, with a negligible bare-molecular component.

In conclusion, we see that the effects of pair-correlations play a major role in atom-molecule coupling, resulting in the appearance of a $1/r$ singularity in the relative wavefunction together with a corresponding decrease in the bare-molecule population. This suggests that for molecule formation it is best to have a weak coupling, while for manipulation of atomic interactions, e.g. for BCS pairing of fermions \cite{HolKokChi01,RegGreJin04}, a strong coupling will remove the corresponding bare-molecule population. In FR the free-space coupling strength is predetermined by atomic properties, hence $\Omega$ can only be increased by decreasing the trap size. In PA, however, the coupling strength is readily increased by increasing the laser intensity. This suggests that laser-induced photoassociation may have a significant advantage over Feshbach Resonance for tuning atom-atom interactions.

\end{document}